\documentclass[a4paper,11pt]{article}
\pdfoutput=1 

\usepackage{jheppub} 

\usepackage[T1]{fontenc} 

\usepackage{amssymb,amsmath,mathbbol,mathrsfs}
\usepackage[usenames,dvipsnames]{color}
\usepackage{hyperref}
\usepackage{stmaryrd}
\usepackage{slashed}

\usepackage[utf8]{inputenc}

\newcommand{\be}{\begin{equation}}
\newcommand{\ee}{\end{equation}}

\newcommand{\Lie}{\mathrm{Lie}}

\renewcommand{\d}{{\mathrm{d}}}
\newcommand{\D}{{\mathrm{D}}}
\newcommand{\DD}{{\mathscr{D}}}

\newcommand{\SU}{{\mathrm{SU}}}

\newcommand{\G}{{\mathcal{G}}}

\newcommand{\pp}{{\partial}}

\renewcommand{\bar}{\overline}
\newcommand{\dd}{\delta}

\renewcommand{\a}{\mathrm{a}}
\newcommand{\scri}{{\mathscr{I}}}
\renewcommand{\i}{{\mathrm{i}}}
\newcommand{\f}{{\mathrm{f}}}
\newcommand{\Q}{Q}

\newcommand{\lbr}{\llbracket}
\newcommand{\rbr}{\rrbracket}

\newcommand{\bb}{\mathbb}

\newcommand{\tr}{\mathrm{Tr}}

\renewcommand{\#}{\sharp}

\title{\boldmath Soft charges from the geometry of field space}

 \author{Aldo Riello}
\affiliation{Perimeter Institute for Theoretical Physics,\\ 31 Caroline St. N., Waterloo, ON N2L2Y5, Canada}

\emailAdd{ariello@perimeterinstitute.ca}

\abstract{Infinite sets of asymptotic soft-charges were recently shown to be related to new symmetries of the $S$-matrix, spurring a large amount of research on this and related questions.
Notwithstanding, the raison-d'être of these soft-charges rests on less firm ground, insofar as their known derivations through generalized Noether procedures tend to rely on the fixing of (gauge-breaking) boundary conditions rather than on manifestly gauge-invariant computations.
In this article, we show that a geometrical framework anchored in the space of field configurations singles out the known leading-order soft charges in gauge theories.
Our framework unifies the treatment of finite and infinite regions, and thus it explains why the infinite enhancement of the symmetry group is a property of asymptotic null infinity and should not be expected to hold within finite regions, where at most a finite number of physical charges -- corresponding to the reducibility parameters of the quasi-local field configuration -- is singled out.
As a bonus, our formalism also suggests a simple proposal for the origin of magnetic-type charges at asymptotic infinity based on spacetime (Lorentz) covariance rather than electromagnetic duality.}

\begin{document} 
\maketitle
\flushbottom


%
%


The work of Strominger and collaborators on the ``soft-triangle'' (\cite{StromLectures} and references therein) unveiled fascinating and unexpected relations among 
(\textit{i}) Weinberg's soft theorems and their generalizations \cite{Weinberg,Low,Cachazo}, 
(\textit{ii}) the so-called (nonlinear) memory effects \cite{Christodoulou,Bieri,Zhiboedov}, and
(\textit{iii}) an enlarged group of asymptotic symmetries for the $S$-matrix, in both gauge theories and gravity.
Strong evidence for these enlarged symmetries has been gathered from the study of the soft theorems. 

However, while descriptions in the asymptotic phase space exist (e.g. \cite{StromLectures, AshtekarStreubel, CampigliaLaddha, CampigliaLaddhaSub, Seraj}), a derivation of these charges from first principles is still missing.
In particular, the relationship between what is ``gauge'' and what is ``symmetry'' is most often obscured by the employment of gauge-fixings that happen to leave only the sought-after {\it symmetry} parameters unfixed.
In other words, the a-priori justification for the existence of these symmetries seems to often rely at some step on the fixing of appropriate (gauge-breaking) boundary conditions. 
This often leads to the question of whether even larger group of symmetries are waiting to be discovered by lifting {\it all} gauge fixing (see e.g. \cite{Donnelly}).
Moreover, to interpret the soft modes as Goldstone modes \cite{StromLectures}, the broken symmetries must be global and not gauge in nature. 
Hence the questions: can the distinction between gauge and global symmetries be made a priori? 
Does this distinction lead to the correct charges?\footnote{Curiously, it turns out that these issues are better understood in gravity \cite{AshtCampLaddha}.}

(For clarity: by symmetry-related configurations, we mean distinct but otherwise degenerate configurations; conversely, by gauge-related configurations, we mean fully degenerate configurations that should be identified within a ``reduced" physical phase space.)

%
%
Early on, the suspicion emerged that these questions are related to the interplay between gauge and the (infinitely far) boundaries\footnote{However, see \cite{Mohd}, who implicitly suggests that the new symmetries have more to do with the null-ness of $\scri$ than with gauge.} of either the spacelike Cauchy surfaces or $\scri^\pm$.
However, given the subtle interplay between boundaries and gauge, and lacking a rigorous distinction between symmetries and gauge, one can easily be led to either (\textit{a}) define gauge too weakly, thus finding spurious boundary charges, or (\textit{b}) too strongly, thus missing relevant ones.

Case (\textit{a}) arises e.g. when gauge transformations are {\it by definition} required to be trivial at the boundary; this is often done to enforce the functional differentiability of the smeared constraints as gauge generators (i.e. as charges). Then, would-be-gauge transformations that are not boundary-trivial inherit a fortiori the status of a symmetry  -- modulo pure-bulk gauge transformations. Since charges are given by smeared constraints only for bulk gauge transformations, one would be led to interpret the ensuing non-vanishing boundary charges as physical, on a par with an angular momentum charge. Although mathematically consistent, we find this viewpoint physically questionable,\footnote{In particular in relation to the dynamical conservation properties usually implicit in the notion of charges.} and indeed not supported by our analysis below (see also \cite[Sect.1]{Hopfmuller}). 

Case (\textit{b}) arises e.g. if from the onset one (naively) treats as gauge {\it all} gauge-like transformations at asymptotic null infinity, even those at its boundary. Then, all configurations compatible with the fall-off conditions could be gauge-fixed so that {\it all} the components of the leading-order gauge potential vanish -- thus disallowing the existence of soft-charges as symmetry generators.

The physically relevant soft-symmetry group relies on a definition of gauge which lies somewhere in between (\textit{a}) ``all boundaries carry symmetries,'' and (\textit{b}) ``no boundary carries symmetries;'' whereas the analogue of attitude (\textit{b}) appears to be valid in finite regions. The goal of this work is to present a mathematical framework capable of providing a unifying definition of gauge and symmetry that strikes this subtle balance both at finite and asymptotic boundaries. 

The relevant framework was developed for finite boundaries in \cite{varpiObs,varpiPRD,varpiYM, GomesRielloWIP}, and will be here developed in the context of asymptotic infinity. As we shall show, (\textit{i}) it is manifestly gauge invariant at boundaries, (\textit{ii}) provides a criterion to distinguish gauge from symmetry, (\textit{iii}) can be applied to infinite Cauchy surfaces as well as to finite bounded regions, (\textit{iv}) associates nonvanishing charges to symmetries but not to gauge, (\textit{v}) engenders an infinite dimensional group of symmetries at $\scri$ {\it only}, (\textit{vi}) associates to these symmetries charges that reproduce the leading (electric) soft theorems, and (\textit{vii}) does not require the addition of new degrees of freedom. 
In other words, our formalism provides at the same time a gauge invariant approach to symplectic geometry in bounded systems, whether at finite or asymptotic distances, and an a priori derivation of the enhancement of symmetries at asymptotic infinity.

As byproducts we will also uncover a simple proposal for the origin of magnetic-type charges at asymptotic boundaries from spacetime covariance, as well as the relation between the quasi-local analogue of the radiative degrees of freedom of \cite{GomesRielloWIP} and the asymptotic Ashtekar-Streubel radiative phase-space \cite{AshtekarStreubel}. 
Notably, we recover the Ashtekar-Streubel phase space through a formalism that applies both to finite and infinite systems, and we do so within a framework that offers extra structures capable of automatically distinguishing ``gauge'' from ``symmetry.''

The techniques employed are geometric in nature, although the investigated geometry is that of field space.
We will focus on Yang-Mills (YM) theories in 4 spacetime dimensions.
In an appendix \ref{app:generalizations}, we comment on generalizations to higher dimensions.

The strategy we follow proceeds in two steps. 
First, we exploit the fiducial fibre bundle structure and the kinematic supermetric on the configuration space of the YM theory to geometrically construct a gauge-invariant quasi-local symplectic potential.
Through this gauge-invariant symplectic potential, we can successfully parse physical symmetries from gauge transformations, and devise a gauge-invariant Noether procedure which assigns nonvanishing charges to the former only.    

On (portions of) the finite-time Cauchy surfaces, our gauge-invariant symplectic form carries only the quasi-local generalization of the gluons' radiative modes, and leaves aside the constrained Coloumbic components of the gauge field which are fully determined by the Gauss constraint.
Only a finite number of gauge-invariant global symmetries and charges is nonvanishing; they arise in the presence of a nontrivial stabilizer (aka reducibility parameter) of the background field configuration. 

Far into the future, close to future null infinity, application of our techniques unveils a symmetry enhancement whose associated leading order charges coincide with the leading order (electric and magnetic) soft charges.
At asymptotic null infinity, two circumstances are fundamental: the induced metric on the hypersurface becomes (conformally) null and the gauge field is assumed to admit an expansion in inverse powers of the radial coordinate.\footnote{Note that this assumption might clash with certain gauge-fixings: e.g. to fix radial gauge $A_r=0$, and thus $A_r^{(1)}=0$, a $\ln(r)$-gauge transformation is needed which undermines the expansion itself.\label{fn:gaugeinv}}
The latter circumstance, related to the peeling property, is a foundational component of the asymptotic-infinity formulation of isolated systems (see e.g. \cite{PenroseBook}).

In our framework, no gauge-fixing or restriction on the gauge parameters -- apart from the requirement that they admit an asymptotic expansion\footnote{The work of \cite{CampigliaLaddha,CampigliaLaddhaSub} suggests that allowing more general asymptotic expansion is a way to recover the subleading soft theorem. In this work, we limit ourselves to the natural asymptotic expansion and leading soft theorems.} -- is ever required.
The origin of the new soft charges is not, as it is sometimes suggested (e.g. \cite[Sect.1]{Hopfmuller}), to be found in the generic breaking of gauge symmetry at the boundary of a Cauchy surface, but rather in the specific geometric and peeling properties associated with asymptotic infinity.
Ensuring manifest gauge-invariance both in the bulk and at the boundary is the cornerstone of the present approach.

\section{Configuration Space Geometry\label{sec:YMgeometry}}

Let $\{\Sigma_t\}_t$ be a Cauchy foliation of a $(3+1)$-dimensional spacetime $M\cong\Sigma_t \times \mathbb R$. 
For notational simplicity, we restrict our analysis to foliations of unit lapse $N=1$ (cf. appendix \ref{app:N}).
However, we keep the shift $\beta_i$ nontrivial, since it will be relevant in the retarded-time coordinates needed for the asymptotic limit.
Our focus is the gauge-variant configuration space of the YM theory with compact semi-simple charge group $G$ associated to $\{\Sigma_t\}_t$.
More precisely, we consider the (sub)configuration space $\Phi=\{A\in\Omega^1(R,\Lie(G))\}$ associated to a finite bounded region $R\subset \Sigma$ such that $R \cong  B_3$ (the $3$-dimensional ball) with $\pp R \cong S_{2}$. 
Crucially, we will not demand gauge transformation to trivialize at $\pp R$.
This is because we are interested in identifying the quasi-local gauge-invariant degrees of freedom, and this cannot depend on a choice of gauge-fixing at $\pp R$, especially when $R$ is a fiducial subregion of $\Sigma$ \cite{varpiObs,varpiPRD,varpiYM,GomesRielloWIP,Henriquephilo}.  

We will denote by $n$ the future-pointing unit timelike normal to $\Sigma$ in $M$, and by $s$ the spacelike outgoing unit normal to $\pp R$ in $\Sigma$.

The configuration space $\Phi$ has the structure of a fiducial, infinite-dimensional, principal fibre bundle $\pi: \Phi \to \Phi/\G, \; A\mapsto [A]$. Here, $\G \cong \mathcal C^\infty(R, G)$, equipped with point-wise multiplication in $G$, is the group of gauge transformations, or ``gauge group'', and $\Phi/\G$ is the space of ``physical configurations'' \cite{Singer,BabelonFP,Babelon,Mitter,DeWittBook}.
It is natural to provide $\Phi$ with two geometric structures. 
The first, associated to $\Phi$'s fibre bundle structure, is a functional connection-form (`\textsc{Var-Pie}'),
\be
\varpi \in \Omega^1(\Phi, \Lie(\G)).
\ee
The second is the ultralocal supermetric that can be read off the kinetic term of the YM Lagrangian, $L = \tfrac12\bb G(\pp_t A, \pp_t A) + \dots$:
\be
\bb G(\delta_1 A, \delta_2 A) = \int_R \d vol \; g^{ij} \tr( \delta_1A_i \delta_2A_j)
\ee
where $g_{ij}$ is the induced Riemannian metric on $\Sigma_t$ (henceforth, we will omit the measure $\d vol = \sqrt{g}\d^D x$).\\

\paragraph{The functional connection}
By construction, $\varpi$ tells the pure-gauge variations of $A$ from the physical ones, in a gauge-covariant manner. 
Let $\xi^\# \in \mathfrak{X}^1(\Phi)$ be the field-space vector associated to an infinitesimal gauge transformation $\xi \in \Lie(\G)$. With a mild abuse of notation,\footnote{More precisely: $\xi_A^\#= \int_R \d vol(x)\;(\D \xi)_i^\alpha(x) \frac{\delta}{\delta A_i^\alpha(x)} \in \mathrm T_A \Phi.$} 
\be
\xi^\#_A = \delta_\xi A \equiv \D \xi \equiv \d \xi + [A,\xi] \in\mathrm{T}_A\Phi.
\label{eq:xihash}
\ee
Then, the defining properties of $\varpi$ can be expressed as
\be
\begin{cases}
\varpi(\xi^\#) = \xi\\
\bb L_{\xi^\#} \varpi = \dd \xi + [\varpi,\xi]
\end{cases}
\label{eq_varpiproperties}
\ee
where $\bb L$ is the field-space Lie-derivative and $\dd$ the (fiducial) field-space de Rham differential. 
To avoid confusion between field-space and spacetime quantities, we reserved double-struck symbols for field-space. 
In the second equation, the bracket $[ \cdot ,\cdot]$ is the Lie bracket in $\Lie(G)$.
Hence, this equation states that $\varpi$ transforms covariantly when transported along the gauge orbits in $\Phi$ (if $\xi$ is a field-{\it in}dependent gauge transformation, then $\dd \xi \equiv0$, see \cite{varpiObs,varpiPRD,varpiYM}).
The first condition states that $\varpi$ is essentially a projector on the tangent space of the gauge orbit, $V_A = \mathrm{Span}(\xi^\#_A)\subset\mathrm T_A\Phi$, and therefore its kernel defines the physical, or ``horizontal'', directions $H_A$. Given a generic variation $\delta A$, its horizontal component is
\be
\delta_H A = \delta A -\varpi^\#(\delta A) = \delta A - \D \varpi(\delta A) \in H_A.
\label{eq:deltaHA}
\ee
Contrary to $\delta A$, $\delta_HA$ transforms covariantly even under field-dependent gauge transfomations ($\delta g\neq 0$), i.e. $\delta_H (A^g) = g^{-1}( \delta_H A )g $.
Notice that although a (local) gauge fixing $\sigma: \Phi/\G \hookrightarrow\Phi$ defines a unique $\varpi$ such that $\mathrm{Im}\mathrm T\sigma\subset H$, the converse is not true: first, because the horizontal distribution $H$ might not be Frobenious-integrable (if $\bb F = \delta \varpi + \varpi^2 \neq 0$), and second, because $\varpi$ is defined along the entire gauge orbit and thus cannot select any \textit{one} section of $\Phi$.   
At this point it is important to notice that $\varpi$ is not uniquely defined, since the algebraic split $\mathrm T\Phi = V \oplus H_\varpi$ is not canonical. 
This leads us to the next ingredient.

\paragraph{The Singer-DeWitt connection}
If $\Phi$ is equipped with a supermetric which is constant in the gauge directions $\bb L_{\xi^\#} \bb G = 0$, the orthogonal decomposition $\mathrm T\Phi =  V \perp H_\perp $ defines a functional connection $\varpi_\perp$ \cite{Singer,DeWitt1967,BabelonFP,Babelon,Mitter,DeWittBook,varpiPRD,varpiYM}, through the condition that $\delta_{H} A$ \eqref{eq:deltaHA} is orthogonal to all vertical vectors $\xi^\#$ \eqref{eq:xihash},
\be
0\stackrel{!}{=} \bb G( \delta_\xi A , \delta A - \D\varpi_\perp(\delta A) ) \quad \forall \xi.
\ee
This readily leads to the elliptic boundary value problem
\be
\begin{cases}
\D^2 \varpi_\perp =  \D^i \delta A_i & \text{in $R$} \\
\D_s \varpi_\perp = \delta A_s & \text{at $\pp R$}
\end{cases}
\label{eq_SdW}
\ee
where the $s$ subindex stands for the contraction with the spacelike normal $s^i$ and $\D_i$ is the gauge covariant generalization of the Levi-Civita (LC) derivative on $\Sigma$.
We will refer to the above connection as the Singer-DeWitt (SdW) connection.
The boundary condition crucially follows from the fact that gauge transformations have not been trivialized at $\pp R$.
In electromagnetism on a flat hypersurface, the above is a Poisson equation with Neumann boundary conditions.
Its solutions are unique up to a constant. 
This remark is the seed of some fundamental considerations that we postpone to the next section on symmetry and charges.

In non-Abelian theories, the SdW curvature $\bb F_\perp = \delta \omega_\perp + \omega_\perp^2$ does not vanish \cite{Singer} and indeed it satisfies the following boundary value problem \cite{varpiYM,GomesRielloWIP}:
\be
\begin{cases} 
\D^2 \bb F_\perp = g^{ij}[\dd_\perp A_i \stackrel{\curlywedge}{,} \delta_\perp A_j] & \text{in }R\\ 
\D_s \bb F_\perp = 0 & \text{at }\pp R .
\end{cases}
\label{eq_bbF}
\ee

We conclude by stressing the fact that \eqref{eq_SdW} imposes boundary conditions on $\varpi$, and not on $\delta A$, which is free to take any value at $\pp R$ or elsewhere. In particular, no boundary condition is imposed on the gauge field.

\section{Quasi-local degrees of freedom\label{sec:sym}}

\paragraph{SdW-horizontal symplectic potential}
The physical relevance of the SdW connection descends from the fact that the supermetric $\bb G$ features in the kinetic term of the configuration-space Lagrangian $L= T-U$, i.e. $T = \tfrac12 \bb G( E, E) = \tfrac12 \bb G(\pp_t A, \pp_t A)+\dots$, where we introduced the electric field $E_i $,
\be
E_i \equiv F_{ni}[A] = \dot A_i  - \D_i A_t,
\label{eq:EFni}
\ee
and the symbol $\dot A_i $,
\be
\dot A_i \equiv \pp_t A_i -  \beta^k F_{ki}
\label{eq:Adot}
\ee
(recall that $\beta_i$ is the shift of the foliation $\{\Sigma_t\}_t$).
In this ``3+1'' decomposition, $A_t$ is a scalar on $\Sigma_t$.
Inspection of the YM action shows that $A_t$ is a Lagrange multiplier for the Gauss constraint, 
\be
\mathcal C_{\mathrm G} = \D^i E_i = \D^i \dot A_i -\D^2 A_t  \approx 0
\ee
(later we will argue that the Gauss constraint within a bounded region $R$ should be complemented by an appropriate boundary condition).
Also, from these equations, it follows that the symplectic potential of a YM theory, when written in configuration space variables (i.e. in terms of the velocity fields, instead of the momenta) and restricted to $R$, reads
\be
\theta = \int_R g^{ij} \tr( E_i \delta A_j )= \bb G( E , \delta A).
\ee

Now, the definition of the horizontal variations $\delta_H A$ allows us to introduce a decomposition of $\theta$ into its horizontal and vertical components:
\be
\theta = \theta_H + \theta_V.
\ee 
The horizontal component $\theta_H$ is obtained by simply replacing $\delta A$ with its horizontal counterpart, while the vertical component is defined by the difference $\theta_V=\theta - \theta_H$:
\begin{align}
\theta_H(A,\delta A) & = \bb G(E , \delta_H A)\\
\theta_V(A,\delta A) & = \bb G(E, \D \varpi)
\label{eq_thetaH}
\end{align} 

Notice that in these formulas, we are interpreting the electric field $E$ as a vector field in $\Phi$.
This is possible because -- forgetting $\D_i A_t$ for now -- $E_i$ is essentially equal to $\pp_t A_i$ (albeit corrected for a nontrivial shift of $\{\Sigma_t\}_t$), and $\pp_t A_i$ can be interpreted as the tangent ``velocity vector'' to the history of $A_i$ in configuration space.
With this interpretation, $\Pi = \bb G(E, \cdot) \in \mathrm T^*\Phi$ is the momentum of $A_i$.  
Decomposing $E$, understood as a vector, into its horizontal and vertical components, we find 
\be
E_i = \dot A_i - \D_i\varpi(\dot A)  + \D_i \varphi =: \dot A_i^H + \D_i \varphi,
\ee
for some $\varphi \in C^\infty(\Sigma,\Lie(G))$ that transforms gauge-covariantly: $\delta_\xi \varphi = [\varphi,\xi]$. 
This decomposition corresponds to writing $A_t$ as
\be
A_t =  -\varphi + \varpi(\dot A),
\label{eq_At}
\ee
which is particularly convenient, since $\varpi(\dot A)$, under a gauge transformation $\xi$, can be seen to transform as $\delta_\xi \varpi(\dot A) = \bb L_{\xi^\#} \varpi(\dot A) = [\varpi(\dot A), \xi] + \pp_t \xi$. 
This uses \eqref{eq_varpiproperties} and the fact that $\dot A$ is a vector on $\Phi$ and thus $\bb L_{\xi^\#}\dot A = \lbr \xi^\#, \dot A\rbr_{\mathrm T\Phi}$ is a Lie bracket between vectors. The formula holds also for field-dependent gauge transformations, provided $\pp_t$ is interpreted as a total derivative $\d/\d t = \pp/\pp_t + \int (\pp_t A) \delta /\delta A$ (for details see \cite[Appendix A]{GomesRielloWIP}). 
Therefore, \eqref{eq_At} automatically takes care of the nontrivial transformation properties that the Lagrange multipliers must satisfy to have full consistency under time-dependent gauge transformations, the inhomogeneous piece being taken care of by $\varpi(\dot A)$, which depends only on the configuration variables.

These considerations hold for any choice of $\varpi$. 
But, for the SdW connection, $\varpi=\varpi_\perp$ and $\varphi=\varphi_\perp$, there is more: this is the {\it only} functional connection such that (\textit{i}) the Gauss constraint $\mathcal C_\mathrm{G}$ depends only on $\varphi=\varphi_\perp$, and (\textit{ii}) $\varphi=\varphi_\perp$ drops from the horizontal symplectic potential. 
These two facts are closely related: they mirror the two steps of symplectic reduction, where one first solves the constraint and then removes the ``conjugate'' degree of freedom.
These two facts also imply that, for the SdW connection, $\varphi=\varphi_\perp$ acquires the interpretation of Coulombic potential, whereas $\theta_H=\theta_\perp$ of the symplectic form for the radiative dof (see also \cite{GomesRielloWIP} for a thorough discussion including $\theta_V$).

Thus, using the SdW decomposition, the Gauss constraint $\mathcal C_\mathrm G = \D^i E_i = \D^2\varphi_\perp$ becomes a Laplace equation for $\varphi_\perp$, that -- in a bounded region -- should be complemented by the appropriate boundary condition, that is
\be\text{Gauss (in vacuum): }
\begin{cases}
\D^2 \varphi_\perp  \approx 0 & \text{in } R\\
\D_s \varphi_\perp \approx f & \text{at } \pp R
\end{cases}
\label{eq_Gauss}
\ee
Both equations are derived using the fact that SdW-horizontal vectors $\dot A_i^\perp$ are by definition in the kernel of $\varpi_\perp$ and therefore annihilate the rhs of \eqref{eq_SdW}. For the boundary condition, we introduced the symbol $f= E_s $ for the flux of $E$ through $\pp R$: this quantity plays the role of an independent datum whose role is to encode the physics from the complement of $R$ in $\Sigma$ (see \cite{GomesRielloWIP} for an in-depth discussion).

From the above, it follows that the symplectic structure gets SdW-decomposed into the following horizontal (radiative) and vertical (Coulombic) components:
\begin{align}
\theta_\perp &= \bb G( E, \delta_\perp A) = \bb G( \dot A^\perp, \delta_\perp A) = \int_R g^{ij} \tr(\dot A^\perp_i\delta_\perp A_j )
\label{eq_thetaperp} \\
 \theta_V &  =  \bb G( E, \D\varpi_\perp) = \bb G( \D\varphi_\perp , \D \varpi_\perp) \approx \oint_{\pp R} \tr(f \varpi_\perp)
\end{align}
These equations are derived from \eqref{eq_thetaH} and the definition of the SdW connection $\varpi_\perp$ as the one for which horizontality is given by $\bb G$-orthogonality to the fibres $\pi^{-1}([A]) \subset \Phi$, as well as from \eqref{eq_Gauss}.

Notice that the first-order perturbations $a_i = \delta_\perp A_i$ can be (loosely) thought as gauge-fixed photons and gluons in a gauge that makes them (covariantly) transverse, $\D^i a_i = 0$. Moreover, the bulk transversality prescription is complemented with adequate and unambiguous boundary conditions, $a_s = 0$.

\paragraph{Quasi-local degrees of freedom} 
To recapitulate, we used the kinetic supermetric $\bb G$ to define a preferred functional connection $\varpi=\varpi_\perp$, named after Singer and DeWitt (SdW), according to the slogan ``horizontality from fibre orthogonality''. 
The preferred status of $\varpi_\perp$ results from the fact that the SdW connection is the only functional connection for which the non-dynamical component of $E$, i.e. the one fixed by the Gauss constraint, drops from $\theta_H=\theta_\perp$.  
In other words, $\varpi_\perp$ is the only choice of functional connection such that the corresponding gauge-invariant (horizontal) symplectic potential $\theta_H=\theta_\perp$ contains only the quasi-local analogue of the radiative degrees of freedom, with no Coulombic contribution left. 
This is possible because of the role the kinetic supermetric $\bb G$ plays in the (usual) symplectic form $\theta$ (when expressed in configuration space).

Notice that the quasi-local radiative degrees of freedom, i.e. the SdW-horizontal perturbations $\delta_\perp A$ are nonlocally built from a generic perturbation $\delta A$, since this requires solving \eqref{eq_SdW}.
Thanks to the presence of boundary conditions in \eqref{eq_SdW}, the nonlocality is limited to the region $R$ and no further information from the rest of $\Sigma$ is required. 
Finally, none of these considerations is affected by the presence of matter. 

E.g., for $G=\SU(N)$, a Dirac fermion $\psi \in \mathcal C^\infty(\Sigma,\bb C^4\otimes W)$, with $W\cong\mathbb C^N$ the fundamental representation of $G$, can be incorporated as follows both into the Gauss constraint,
\be
\text{Gauss: }
\begin{cases}
\D^2 \varphi_\perp  \approx \rho & \text{in } R\\
\D_s \varphi_\perp \approx f & \text{at } \pp R,
\end{cases}
\label{eq_Gaussmatter}
\ee
and into the SdW decomposition of $\theta$,
\begin{align}
\theta_\perp& = \int_R g^{ij} \tr(\dot A^\perp_i \delta_\perp A_j )  - \tfrac{1}{2}(\bar\psi \gamma^0 \dd_\perp \psi - \delta_\perp \bar\psi \gamma^0 \psi) 
\label{eq_thetaperpmatt}\\
\theta_V & \approx \oint_{\pp R} \tr( f \varpi_\perp)
\end{align}
where $\bar\psi = \psi^\dagger\gamma^0$ is the adjoint spinor; the charge density $\rho$ is defined by $\tr(\rho\xi) =  \bar\psi \gamma^0 \xi \psi $ for all $\xi\in\Lie(G)$; and the matter horizontal vectors read $\delta_\perp \psi = \delta \psi + \varpi_\perp \psi$ and $\delta_\perp\bar\psi = \delta\bar\psi - \bar\psi \varpi_\perp$, respectively.\footnote{Conventions are as in \cite{GomesRielloWIP}.}
Crucially, $\varpi_\perp$ is still defined purely in terms of $\delta A$ according to \eqref{eq_SdW}.
In electromagnetism, $\delta_\perp \psi$ and $\delta_\perp A_i$ can be interpreted in terms of quasi-local generalization of the variations of a Dirac-dressed scalar field and of transverse photons, respectively.
However, these interpretations are not fully satisfactory. 
They carry the same benefits and drawbacks of an interpretation of $\varpi$ as a gauge-fixing: although it might provide an intuition, it captures only a limited aspect of the formalism and therefore is, in the end, inadequate \cite{varpiYM}. 

We conclude this section with the SdW decomposition of the symplectic form $\Omega = \dd \theta$ on-shell of the Gauss constraint, i.e. $\Omega \approx \Omega_\perp + \Omega_\pp$. This decomposition is defined by isolating the horizontal term (the last equality follows from the gauge invariance of $\theta_\perp$)
\be
\Omega_\perp \equiv \dd \theta_\perp = \dd_\perp\theta_\perp,
\ee
and reads
 \begin{align}
\Omega_\perp & = \int_R  g^{ij} \tr\Big( ( \pounds_n \dd_\perp A_i ) \curlywedge \dd_\perp A_j \Big)  - \delta_\perp {\bar\psi} \curlywedge \gamma^0 \delta_\perp \psi - \tr(\rho \bb F_\perp)
\label{eq_Omegaperp}\\
\Omega_\pp & = \oint_{\pp R} \tr(\dd_\perp f \curlywedge \varpi + f \bb F_\perp)
\end{align}
where $\pounds_n$ is a {\it gauge-covariant} version of the Lie-derivative along the normal to the foliation $\{\Sigma_t\}_t$, $n = \pp_t - \beta^i\pp_i$:
\be
\pounds_n \dd_\perp A_i  
\equiv \widetilde\D_t \dd_\perp A_i  - \beta^k \D_k \dd_\perp A_i - (\nabla_i \beta^k) \dd_\perp A_k . 
\label{eq_Ln}
\ee
Here, $\widetilde\D_t \equiv \pp_t + [\varpi(\dot A), \cdot\,]$ is a gauge covariant derivative just as $\D_t = \pp_t + [A_t ,\cdot\,]$ (Indeed, these two derivatives differ by a term involving the Coulombic potential, which transforms in the adjoint representation, cf. \eqref{eq_At}).
In a field-space analogy to the wedge product of differential forms on spacetime, ``$\curlywedge$'' denotes the antisymmetric product of field variations.

As already emphasized, the first order perturbation $a_i = \delta_\perp A_i$ can be loosely understood as gauge-fixed (covariantly) transverse photons or gluons on a generic background $A$.
In this sense, $\Omega_\perp$ is a finite-region and manifestly gauge-covariant analogue of the asymptotic Ashtekar-Streubel radiative phase-space structure (more on this later). 
We will henceforth refer to $\Omega_\perp$ as the ``radiative'' symplectic form.

Whereas the expression of $\Omega_\pp = \dd \theta_V$ readily follows from the identities $\dd f = \dd_\perp f +[f,\varpi_\perp]$ and $\dd \varpi_\perp  = \bb F_\perp  - \tfrac12[\varpi_\perp\stackrel{\curlywedge}{,}\varpi_\perp]$, that of $\Omega_\perp$ is much subtler to derive -- see app. \ref{app:Xirad}.

\section{Symmetries and charges\label{sec:sym}}

As remarked above, in flat space electromagnetism, the defining equation of the SdW connection \eqref{eq_SdW} admits unique solutions only up to constant offsets.
This is an important observation, since $\chi=cte$ corresponds to elements in $\Lie(\G)$ such that $\delta_\chi A =0$. 
This generalizes to non-Abelian YM theories as follows: at $A\in\Phi$, the SdW connection is defined up to reducibility parameters of $A$, i.e. up to a $\chi\in\Lie(G)$ such that $\delta_\chi A=0$. 
Reducibility parameters are the analogue of Killing vector fields in general relativity, and exist only at peculiar, reducible, configurations in $\Phi$.
The geometrical reason for a kernel in the above equations is that $\Phi$ fails to be a bona-fide principal fibre bundle, because certain fibres (i.e. gauge orbits) degenerate, making $\Phi/\G$ a stratified manifold.
The lower the stratum of $ \Phi/\G$ to which $[A] $ belongs, the more symmetric $[A]$ is \cite{varpiYM}.

On a compact and boundary-less hypersurface $\Sigma$, the Hamiltonian generator of the YM gauge transformation $\xi$ is $\Q[\xi] = \theta(\xi^\#) = \theta(A,\psi ;\delta_\xi A,\delta_\xi \psi)$ and vanishes on-shell of the Gauss constraint, $\Q[\xi]\approx0$.
This means that gauge transformations have trivial charges. 
In presence of boundaries, however, $\Q[\xi] \approx q[\xi] $ for $q[\xi] = \oint_{\pp R} \tr(\xi f) \neq 0$. 
This poses a puzzle: do gauge transformation at the boundary of a fiducial region $R\subset\Sigma$ suddenly turn into physical symmetries carrying nontrivial charges? This sounds implausible, unless (possibly gauge breaking) boundary conditions are imposed at $\pp R$. 
This puzzle is resolved if one makes use of the ``horizontal symplectic geometry'' \cite{varpiObs, varpiPRD, varpiYM, GomesRielloWIP}.
Indeed, the contraction of a vertical vector $\xi^\#$ with the horizontal form $\theta_\perp$ identically vanishes,
\be
\Q_\perp[\xi] = \theta_\perp(\xi^\#) \equiv 0,
\ee
unless $\xi=\chi$ is a reducibility parameter of $A$, i.e. $\D\chi=0$, in which case the generator coincides with the physical charge density of the matter distribution \cite{varpiPRD,varpiYM,GomesRielloWIP}:
\be
\Q_\perp[\chi] = \theta_\perp(\chi^\#) = - \int_R \tr( \chi \rho) \approx \oint_{\pp R} \tr(\chi f) = q[\chi]. 
\label{eq_killingcharge}
\ee
In contrast to the other equations in this section, \eqref{eq_killingcharge} depends on the use of the SdW-connection since, in this case, $\varpi_\perp(\chi^\#) = 0$. 
This means that a reducible vertical transformation $\chi^\#$ is also horizontal, i.e. ``physical,'' with respect to the SdW connection \cite{varpiObs,varpiPRD,varpiYM,GomesRielloWIP}.
Therefore, we see that another property of the SdW connection is to select the physically meaningful charges associated to actual global symmetries of the gauge field configuration.
Notice that the notion of reducibility parameter for a configuration $A$ is fully gauge-covariant and thus intrinsic to $A$; this is precisely analogous to Killing vector fields being intrinsically related of a metric tensor.

The Hamiltonian flow equation on the radiative phase space characterized by $\Omega_\perp$, is equivalent to the gauge invariance of $\theta_\perp$:
\be
0 = \bb L_{\xi^\#} \theta_\perp = \dd \bb i_{\xi^\#} \theta_\perp + \bb i_{\xi^\#} \dd \theta_\perp = \dd Q_\perp[\xi] + \bb i_{\xi^\#} \Omega_\perp,
\ee
where we used Cartan's formula for the (field-space) Lie-derivative and the fact that $\dd\theta_\perp = \dd_\perp\theta_\perp$.
Notice that the flow equation is nontrivial, i.e. does not read $0=0+0$, if and only if $\xi = \chi$ is a reducibility parameter.\footnote{For the left-most equality to be satisfied, one also has to require that $\delta \chi + [\varpi_\perp , \chi] =0$, which essentially states that reducibility parameters $\chi$ must be field-dependent and transform in the adjoint representation, $\bb L_{\xi^\#} \chi = [\chi, \xi]$. Transformations that are valued in the reducibility parameter but do not have the correct field-space dependence are neither symmetries nor gauge, $\theta_\perp$ is not invariant along their flow.} This observation allows us to unambiguously differentiate between gauge and symmetry transformations, and to assign non-vanishing charges only to the latter -- even in the presence of boundaries.

Notice that, in YM theory, if a continuous family of reducibility parameters $\{\chi(t)\}_t$ exists along the on-shell motion $\gamma = \{A_i(t) \}_t\subset\Phi$, then $\chi(t)$ is a reducibility parameter of the (on-shell) spacetime connection $A_\mu(t)$. To these spacetime reducibility parameters, one associates proper dynamically conserved charges, since then the spacetime divergence $\nabla_\mu\tr( J^\mu \chi)$ vanishes on-shell (see also \cite{Abbott,DeWittBook,BarnichBrandt}). 
Therefore, our configuration space reducibility parameters are natural candidates for dynamically conserved charges (and in the instantaneous configurations space $\Phi$ this is the best one can do).
In the next section, we apply this construction to fields on $\Sigma_{t=+\infty} = \scri^+$.
This construction will feature fundamentally new properties with respect to the finite-time hypersurface case.

\section{Asymptotic infinity\label{sec:asympt}}

Consider Minkowski space in the retarded coordinates $(t,u,y^A)$, with lapse $N=1$ and shift $\beta_i\d x^i = - \d u$,
\be
\d s^2 = - 2 \d t \d u + \d u^2  + (t-u)^2 \gamma_{AB} \d y^A \d y^B,
\ee
and consider a family of regions $R_t\in\Sigma_t$ defined by $u\in[u_\i,u_\f]$.
In the late time limit, $t\to\infty$:  $\Sigma_t \to \scri^+$, $\gamma_{AB}$ becomes the (conformal) metric on the celestial sphere $S\cong S_2$, and the gauge field is assumed to approach the (vacuum) configuration $A^{(0)}$ at a certain rate. 
In particular, the assumption is made that in a neighbourhood of $\scri^+$ the field admits an expansion\footnote{At least up to a certain order $k_o<\infty$.} in powers of $t$ (which at leading order, and at fixed $u$, coincides with Penrose's conformal factor):
\be
A_i(x,t) = A^{(0)}_i + \tfrac1t A^{(1)}_i + \tfrac{1}{t^2} A^{(2)}_i + \dots
\ee
Hence, at late time, configuration space $\Phi$ has an extra structure that gauge transformations must also respect.
Therefore, they also come layered in inverse powers of $t$:
\be
\xi= \xi^{(0)}_i + \tfrac1t \xi^{(1)} + \tfrac{1}{t^2} \xi^{(2)}+ \dots
\label{eq_asymptsyms}
\ee 
Notice that we are not gauge-fixing the vacuum $A^{(0)}$, nor any of the subleading orders of $A$ (at least to the extent to which they admit the above expansion, see footnote \ref{fn:gaugeinv}).
Once again, we will leave it to the configuration space geometry to distinguish gauge from physical symmetries.
Observe that, in the asymptotic limit, \eqref{eq_SdW} becomes 
\be
\begin{cases}
(\D^{(0)}_u)^2\varpi_\perp^{(0)}  = \D^{(0)}_u \delta A^{(0)}_u +O(\tfrac1t) & \text{in $R_{t\to\infty}$}\\
\D^{(0)}_u\varpi^{(0)}_\perp  = \delta A^{(0)}_u +O(\tfrac1t) & \text{at $\pp R_{t\to\infty}$}
\end{cases}
\label{eq_SdWscri}
\ee
Since these leading-order equations are one-dimensional, they are equivalent to $\D^{(0)}_u \varpi_\perp^{(0)} = \delta A_u^{(0)}$ that is 
\be
\delta_\perp^{(0)} A_u^{(0)} = 0\quad\text{in $R_{t\to\infty}$}.
\label{eq_gaugeAu}
\ee
Therefore, in $R_{t\to\infty}\subset\scri^+$, the horizontal modes can be loosely understood as gauge-fixed perturbations $a^{(0)}_i = \dd_\perp A_i^{(0)}$ in the gauge $a^{(0)}_u=0$. As expected, this leaves us with two degrees of freedom per point of $\scri^+$, i.e. $a^{(0)}_A = \delta_\perp A_A^{(0)}$.

Using the fact that physical symmetries $\chi^\#$ are given by transformations which are simultaneously vertical and horizontal, i.e. in the kernel of \eqref{eq_SdWscri}, we immediately find that -- thanks to the degenerate nature of the limit metric $g_{ij}(t)$ and to the layering of $A\in\Phi_{t\to\infty}$ and $\xi\in\Lie(\G)_{t\to\infty}$ -- the leading order $\chi^{(0)}$ does {\it not} need to be a reducibility parameter of $A_i^{(0)}$, as was the case for finite-boundaries: it suffices that $\chi^{(0)}$ is covariantly constant in retarded time:
\be
\D_u^{(0)} \chi^{(0)} = 0 .
\label{eq_Killingscri}
\ee 
Since the space-like Laplacian is missing, this equation gives us a celestial-sphere-worth of asymptotic {\it symmetries} for all configurations in $\Phi_{t\to\infty}$ -- as opposed to just a finite number of them existing at reducible configurations within finite regions.
In particular, $A_A^{(0)}$ and $A_A^{(0)}+ \D_A \chi^{(0)}$ are two (infinitesimally close) backgrounds that are {\it not} gauge-related, but only symmetry-related, and thus -- according to the ``horizontal = physical'' slogan -- should not be identified. This is precisely the enlarged asymptotic symmetry group between non-equivalent backgrounds discussed in \cite{KacMoody}. 
In our framework we see that the enlargement is the consequence of both the assumed fall-offs and of points on the celestial sphere being infinitely separated from each other. 

To compute the asymptotic charge of the physical symmetries $\chi^{(0)}$ through the first equality in \eqref{eq_killingcharge}, we must first define $\theta_{\perp}$ on $\Phi_{t\to\infty}$, verify its finiteness and hence evaluate it at $\delta A = \chi^{(0)\#}$.
Observe that even for $t\to\infty$, $\theta_\perp$ can be written as 
\be
\theta_\perp = \bb G( E, \delta_\perp A) = \bb G(\dot A, \delta_\perp A),
\label{eq_thetaperp2}
\ee
if $A_t$ is assumed to admit an expansion in $t^{-1}$.
This follows from the general equality, $\bb G(E + \D\phi, \delta_\perp A) = \bb G(E,\delta_\perp A)$ for all $\phi$ admitting an expansion in $t^{-1}$, due to the defining relation of $\varpi_\perp$ to $\bb G$.
To proceed, we expand $\dot A_i = \pp_t A_i + F_{ui}$, obtaining in particular
\be
\begin{cases}
\dot A_u =  O(\tfrac{1}{t^2})\\
\dot A_A = F^{(0)}_{uA}+ O(\tfrac1t)
\end{cases}
\ee
Notice that $F^{(0)}_{uA} = \pp_u A_A^{(0)} - \DD_A A_u^{(0)}$, where $\DD_A$ is the gauge-covariant LC derivative of $\gamma_{AB}$ on the celestial sphere $S$, and $A_u^{(0)}$ is a scalar on $S$.
Hence, inserting these results in \eqref{eq_thetaperp2} and using that $\dd_\perp A_u^{(0)}=0$  \eqref{eq_gaugeAu}, we get
\begin{align}
&\theta_\perp  = \int_{\scri^+}  \tr\Big( \gamma^{AB} F_{uA}^{(0)}\delta_\perp A^{(0)}_B \Big) + O(\tfrac1t), 
\label{eq_thetainfty}
\end{align}
where we abbreviated $\int_{\scri^+} = \int_{u_\i}^{u_\f} \d u \int_{S_{(D-1)}} \sqrt{\gamma} \d^{D-1} y$.
Varying this, or from a limit of \eqref{eq_Omegaperp}, we find the following radiative symplectic form on $\scri^+$,
\begin{align}
\Omega_\perp &= \int_{\scri^+} \gamma^{AB} \tr\Big( (\D^{(0)}_u \delta_\perp A_A^{(0)} ) \curlywedge \delta_\perp A^{(0)}_B\Big) + O(\tfrac1t).
\end{align}
 This is nothing else than a gauge-covariant and non-Abelian generalization of the Ashtekar-Streubel symplectic structure of electromagnetism \cite{AshtekarStreubel}. Comparing to \eqref{eq_Omegaperp}, we see that the ``gauge condition'' on the horizontal gluon perturbations has naturally gone from imposing transversality in the bulk, $\D^i\delta_\perp A_i=0$, to imposing temporal gauge on $\scri^+$, $\delta_\perp A_u^{(0)}=0$.

Finally, defining the charge as in the first equality of \eqref{eq_killingcharge} and using \eqref{eq_Killingscri}, we obtain
\begin{align}
\Q_\perp[\chi^{(0)}] = \theta_\perp(\chi^{(0)\#}) = - \int_{\scri^+}  
\tr\Big(\chi^{(0)}\DD^A F_{uA}^{(0)}\Big) + O(\tfrac1t),
\label{eq_Qsoft}
\end{align}
These charges are readily recognized as matching Strominger's ``new asymptotic charges'', or ``soft charges'' \cite{He:2014cra,StromLectures}. 
The term ``soft'' is due to the fact that, in a gauge where $A_u^{(0)}=0$, $\chi^{(0)}$ is constant in time \eqref{eq_Killingscri} and the charge only involves the zero-mode (in retarded time) of the momentum conjugate to the gauge-invariant photon field $\delta_\perp A^{(0)}_A$,
\be
N_A = \int \d u \,F_{uA}^{(0)} \qquad ( \text{in the gauge } A_u^{(0)} = 0 ).
\ee 
(Obvious generalizations involving $u$-ordered exponentials hold for non-gauge-fixed backgrounds.)

The soft charge $Q_\perp[\chi^{(0)}]$ can be re-written in terms of the charge aspect $\sigma  = F_{tu}^{(2)}$ (this is the asymptotic electric field generated by a bulk charge density) by using the asymptotic limit of the Gauss constraint $\D^i E_i \approx 0$ -- we suppose no charge is present in the proximity of $\scri^+$.
At the lowest physically-relevant order in $1/t$, the asymptotic Gauss constraint reads 
\be
\D_u^{(0)} \sigma+ \DD^A F^{(0)}_{uA} \approx 0.
\ee
Using this equation, the definition of the reducibility parameter $\chi^{(0)}$ \eqref{eq_Killingscri}, and explicitly performing the integration in $\d u$, the soft charge \eqref{eq_Qsoft} can be rewritten as 
\be
\Q_\perp[\chi^{(0)}] \approx \oint \tr\Big( \chi^{(0)} ( \sigma(u_\f) - \sigma(u_\i) )\Big).
\ee
Each of these terms is the asymptotic analogue of the charges $q[\chi]$ \eqref{eq_killingcharge} -- except that the set of asymptotic reducibility parameters is much richer. 

\section{Massive charged matter}
Generalization to the presence of massive charged matter requires the resolution of future timelike infinity by the introduction of a spacelike future hyperboloid. For this, one defines $\Sigma_{\tau} $ to be given by the union $\mathcal H^+_\tau \cup \scri^+$, where $\mathcal H^+_\tau = \{ x\in \mathbb R^4 | t^2-r^2 = \tau^2, \, t\geq r\} $ which meets $\scri^+$ at $t=\infty$ \cite{CampigliaLaddha}. 
Now, reducibility parameters that solve the horizontality conditions on $\Sigma_\tau$ are given by a  $\Lie(\G)$-valued function $\Xi: \mathcal H^+_\tau \cup \scri^+ \to \Sigma$ that we write as the combination of a $\Lie(\G)$-valued function $\lambda$ on $\mathcal H^+_\tau$ and $\Lie(\G)$-valued function $\chi^{(0)}$ on $\scri^+$. Continuity of $\Xi$ requires $\lambda = \chi^{(0)} $ at $\pp \mathcal H^+_\tau = \mathcal H^+_\tau \cap \scri^+$. Thus, we find the following horizontality conditions on for $\Xi = (\lambda, \chi^{(0)})$:
on $\scri^+$, $\chi^{(0)}$ must satisfy $\D_u^{(0)}\chi^{(0)}=0$ as before; while, on $\mathcal H^+_\tau$, $\lambda$ should satisfy the boundary value problem:
\be
\begin{cases}
\D^2_\mathcal H \lambda = 0 & \text{on $\mathcal H^+_\tau$}\\
\lambda = \chi^{(0)} & \text{at $\pp \mathcal H^+_\tau = \mathcal H^+_\tau \cap \scri^+$}
\label{eq_lambda}
\end{cases}
\ee
where $\D^2_\mathcal H$ is the gauge-covariant Laplacian on $\mathcal H^+_\tau$.
This follows from the continuity of $\Xi=(\lambda,\chi^{(0)})$ at the interface between $\scri^+$ and $\mathcal H^+$ and from the fact that the only boundary of $\Sigma_\tau$ is the past corner of $\scri^+$.
The fact that boundary conditions are imposed at the past boundary of $\scri^+$ provides the reason why $\lambda$ in \eqref{eq_lambda} does not need to be a reducibility parameter of $A$ on $\mathcal H^+$.\footnote{Reducibility parameters $\chi$ in $R$ are fully characterized by the ``covariant-{\it Neumann}'' boundary value problem:
$$
\begin{cases}
\D^2 \chi = 0 & \text{on $R$}\\
\D_s \chi = 0 & \text{at $\pp R$}
\end{cases}
$$
Notice how the boundary condition for $\chi$ at $\pp R$ is replaced  in \eqref{eq_lambda} by the continuity condition for $\lambda$ at the interface $\pp\mathcal H^+ = \mathcal H^+ \cap \scri^+$. }
The above condition agrees with what was used in \cite{CampigliaLaddha} to generalize the soft charges in the presence of charged massive particles and show their equivalence with the relevant soft-theorems. 
Whereas in \cite{CampigliaLaddha} this equation was derived through the imposition of Lorentz gauge in the bulk, here it is derived using manifestly gauge-invariant techniques intrinsic to $\Sigma_\tau$.

\section{Towards magnetic soft charges}

In \cite{Campiglia:2015qka,CampigliaLaddhaSub}, the relevance of magnetic, rather than electric, soft charges was identified.
The relation of these charges to gauge transformations raises some puzzles, since electro-magnetic duality is explicitly broken -- even in vacuum -- by the introduction of an electromagnetic gauge potential $A_\mu$ subject to gauge transformations.
This is of course related to the fact that since the magnetic charge density identically vanishes, one turns the would-be source equation for the magnetic field into an algebraic identity (Bianchi).
In the present construction, this state of affairs is highlighted by the fact that only the (total) electric charge arises as a horizontal charge on a finite-time hypersurface $\Sigma_t$.
Therefore, the proposed field-space analysis is consistent on finite-time hypersurfaces. 
Again, the question arises, whether it is possible for it to recover magnetic charges asymptotically.

Although we do not here attempt to perform a complete analysis that could fully recover the results of \cite{CampigliaLaddha,CampigliaLaddhaSub} (but see also \cite{CampHopfFrSoni} for a challenging example),
we still like to put forward the following observation: asymptotically magnetic charges can arise from a demand of covariance in the choice of the foliation $\Sigma_t\to\scri$. This is nothing else than a consequence of the usual mixing of electric and magnetic fields under boosts.

Note that whereas the demand for the ``boost covariance'' of charges is a meaningful request at $\scri$, it does not apply on a finite Cauchy surface. This is because, in the bulk, charges can be expressed as integrals over closed $(D-1)$-surfaces $\pp R$, and these surfaces automatically pick a Lorentz frame: within spacetime, $\pp R$ is invariantly described by an area bivector\footnote{In local spacetime coordinates, the surface $\pp R\subset \bb R^{D+1}$ is given by $(\sigma_1,\sigma_2)\mapsto x_{\pp R}^\mu(\sigma_1,\sigma_2)$, and $\mathrm{a}^{\mu\nu}= \pp_1 x_{\pp R}^{[\mu}\pp_2 x_{\pp R}^{\nu]} $.} $\mathrm{a}^{\mu\nu}$ which provides a unique definition of {\it electric} flux: $f=\tfrac12\mathrm{a}^{\mu\nu}F_{\mu\nu}$. Thus, consistently the covariance argument does not lead, at finite boundaries, to the introduction of extra magnetic charges.

Back to $\scri$. Consider a foliation rotating in the axial direction, $\phi \mapsto \phi + \tau t^{-1} $, that comes at leading order in $t^{-1}$ with a shift $\beta_i \d x^i = - \d u - \tau \d \phi +O(t^{-2})$ (recall that $t^{-1}$ is our conformal parameter). 
This shift corresponds to a frame with a finite rotational velocity at infinity. 
This more general shift introduces in the expression of the electric field asymptotically conjugate to $\delta A_A$, a new term, $\tau F_{\phi A}$, see (\ref{eq:EFni},\ref{eq:Adot}). 

In $D+1=4$, this term leads to a new contribution to the soft charge equal to 
\be
\tilde Q[\chi^{(0)}] = \oint \tr\Big(\mathcal B \tau \epsilon_{\phi A} \DD^A \chi^{(0)}\Big) 
\ee
where $\mathcal B = \tfrac12 \int \d u \epsilon^{AB} F^{(0)}_{AB} $ is the (retarded-time) zero-mode of the leading order magnetic flux across the celestial sphere.
This shows that magnetic soft charges do not have to arise necessarily from an electromagnetic duality, but in certain circumstance can be a consequence of a ``covariantization'' of the choice of foliation approaching $\scri$. In other words, asymptotic reference frames that are boosted with respect to each other will see different electric charges that differ from each other by a magnetic charge contribution. This argument explains why one should expect the soft theorems to ``know'' about the magnetic charges too.

This type of soft charge shares properties with the magnetic charges entering Low's subleading soft-theorem \cite{Lysov,CampigliaLaddha,CampigliaLaddhaSub}.
Once again, we did not prove that the present framework is capable of naturally recovering the subleading charges: this would require considerably more work and likely some new ideas. In particular, we did not shed light on the relevance of asymptotically divergent gauge transformations \cite{CampigliaLaddhaSub}. We leave these investigations to future work.

\acknowledgments
It is a pleasure to thank Henrique Gomes, with whom many of the ideas present in the first part of this paper have been developed. A thank goes also to Beatrice Bonga, for her valuable comments on an earlier draft.
Finally, I am indebted with an anonymous referee whose questions and comments have led to a better manuscript -- of course, the responsibility for any mistake or lack of clarity rests only with me.
This research was supported in part by Perimeter Institute for Theoretical Physics. Research at Perimeter Institute is supported by the Government of Canada through the Department of Innovation, Science and Economic Development Canada and by the Province of Ontario through the Ministry of Economic Development, Job Creation and Trade.



\appendix

\section{Generalization to higher dimensions, $D+1>4$ \label{app:generalizations}}

In dimension $D+1>4$, the $t\to\infty$ limit of the symplectic potential \eqref{eq_thetainfty} gives a quantity that a priori diverges as $t^{D-3}$. 
In \cite{FrHopfR}, it is shown how these divergences can be reabsorbed on-shell into the spacetime and field-space cohomological ambiguities of $\theta$. 
The paper deals with electromagnetism (Abelian YM), but work in progress by the authors shows that the same renormalization procedure is possible in general relativity.
There, the conformal factor was chosen to be the more standard inverse radial coordinate $r^{1-}$, rather than Minkowski time, therefore some details of the presentation might undergo slight changes. 
Nonetheless, since the difference in the two choices can be reabsorbed into a retarded-time dependence of $\gamma_{AB}$ ($q_{AB}$ there), the general renormalizability argument of \cite{FrHopfR} will not be compromised, nor we expect the final results to be much different.
Then, thanks to the detailed analysis of \cite{FrHopfR}, we expect the (renormalized) soft charges computed from the limiting renormalized horizontal symplectic potential and $\chi^{(0)}$ as in \eqref{eq_Killingscri} and \eqref{eq_Qsoft} to coincide with those deduced by Strominger and collaborators from the soft theorems \cite{KapecDgeq6}.
All the quoted results hold in even spacetime dimensions larger than $D+1\geq6$.
Electromagnetism and gravity do not admit a conformal compactification in odd spacetime dimensions in presence of radiation (see \cite{Hollands} and \cite{FrHopfR} for a bulk and $\scri$ perspective, respectively).

\section{The radiative symplectic form \label{app:Xirad}}

Define the symplectic structure $\Omega_\perp = \delta \theta_\perp$. From the gauge invariance of $\theta_\perp$ it follows that $\Omega_\perp = \delta_\perp \theta_\perp$ and hence $\Omega_\perp$ is purely horizontal. 

We start by ignoring matter. Using $\Omega_\perp = \delta_\perp \theta_\perp$ and the identity $\delta_\perp(\delta_\perp A_i) = - \D_i \bb F_\perp$, we find 
\be
\Omega_\perp = \int_R g^{ij} \tr( \delta_\perp \dot A_i^\perp \curlywedge \delta_\perp A_j  - \dot A_i^\perp \D_j \bb F_\perp) = \int_R g^{ij} \tr( \delta_\perp \dot A_i^\perp \curlywedge \delta_\perp A_j ),
\label{eq_B1}
\ee
where the SdW curvature term dropped because it is the $\bb G$-inner product between a vertical quantity, $\D_i \bb F$, and and a SdW-horizontal one, $\dot A_i^\perp$.

To put $\delta_\perp \dot A_i^\perp$ in a form that emphasizes the dependence on $\delta_\perp A_i$, we need two identities. The main one,
\be
\dd_\perp \left( (\pp_t A_i)^\perp\right) = \pp_t \dd_\perp A_i + [\varpi(\dot A), \dd_\perp A_i] + \D_i( \cdots ),
\ee
was proven\footnote{In the referenced paper, the proof was performed with the restrictive hypothesis $\pp_t g_{ij} =0$. However, at the expenses of complicating the expression summarized by $(\cdots)$, this hypothesis can be dropped.}  in \cite[App.C]{GomesRielloWIP}. 
The second identity is
\be
\dd_\perp \left( (\beta^kF_{ki})^\perp \right) = \dd_\perp \left( \beta^kF_{ki} - \D_i \varpi_\perp(\beta^kF_{ki} ) ] \right) 
= 2\beta^k\D_{(k} \delta_\perp A_{i)} - [\delta_\perp A_i, \varpi_\perp (\beta^kF_{ki})] + \D_i(\cdots).
\ee
Subtracting these two expressions (recall that $\dot A_i = \pp_t A_i - \beta^k F_{ki}$), we obtain
\be
\delta_\perp (\dot A_i^\perp) = \pp_t \dd_\perp A_i + [\varpi_\perp(\dot A), \dd_\perp A_i] - \beta^k \D_k \dd_\perp A_i + \beta^k \D_i \delta_\perp A_k + \D_i( \cdots ).
\ee
or, readjusting the last term (here $\nabla_i$ is the LC covariant derivative wrt $g_{ij}$):
\be
\delta_\perp (\dot A_i^\perp) = \pp_t \dd_\perp A_i + [\varpi_\perp(\dot A), \dd_\perp A_i] - \beta^k \D_k \dd_\perp A_i -   \delta_\perp A_k \nabla_i \beta^k + \D_i( \cdots ).
\ee
Finally, recalling the notation of \eqref{eq_Ln}, we recognize
\be
\delta_\perp (\dot A_i^\perp) = \pounds_n  \dd_\perp A_i  + \D_i( \cdots ).
\ee

Reintroducing matter into the picture, using $\dd_\perp(\dd_\perp \psi) = \bb F \psi$, and dropping the terms $\D_i(\cdots)$ for the same reason as in \eqref{eq_B1}, we find
\be
\Omega_\perp = \int_R g^{ij}\tr\Big( (\pounds_n  \dd_\perp A_i ) \curlywedge  \delta_\perp A_j \Big) - \dd_\perp \bar\psi \curlywedge \gamma^0 \dd_\perp \psi - \tr(\rho \bb F) 
\label{eq_B7}
\ee
Notice that $\widetilde \D_t = \pp_t + [\varpi(\dot A), \cdot \,]$  differs from $\D_t = \pp_t + [A_t, \cdot\,]$ by the term $[\varpi_\perp, \cdot \,]$. Since the Coulombic potential $\varphi_\perp$ transforms in the adjoint representation, $\widetilde \D_t$ is also a gauge-covariant derivative enjoying all the gauge-covariance properties of $\D_t$. For more details, see \cite[sec. 3.5 \& app. C]{GomesRielloWIP}.

\section{Lapse $N\neq1$\label{app:N}}

Here we report the generalization of some key formulas for the case of lapse $N\neq1$.
The kinetic term of the Lagrangian, $L=T-U$, is $T=\tfrac12\bb G(N E, N E) = \tfrac12 \bb G(\pp_t A, \pp_t A) + \dots$ for the following kinetic supermetric
\be
\bb G(\delta_1 A, \delta_2 A) = \int_R \d vol \; N^{-1} g^{ij}\tr(\delta_1 A_i \delta_2 A_j)
\ee
and electric field
\be
E_i = F_{ni}[A] = \tfrac1N (\dot A_i - \D_i A_t).
\ee
As a consequence of the lapse appearing in the equation above, both the Gauss constraint and the SdW boundary value problem have to be slightly modified according to
\be
N\mathcal C_{\rm G} = \D^i \dot A_i - \a^i \dot A_i - \D^2 A_t + \a^i\D_i A_t \approx 0
\ee
and
\be
\begin{cases}
\D^2 \varpi_\perp - \a^i\D_i\varpi_\perp = \D^i \delta A_i - \a^i\delta A_i & \text{in $R$}\\
\D_s \varpi_\perp = \delta A_s
\end{cases}
\ee
where $\a_i=\D_i\ln N$ is the acceleration of Eulerian observers of $\{\Sigma_t\}_t$.
Similarly, the (pre)symplectic potential, horizontal or not, become: 
\begin{align}
\theta & = \int_R g^{ij} \tr(E_i\delta_\perp A_j) = \bb G(N E,\delta A),\\
\theta_\perp & = \bb G( NE,\delta_\perp A) =  \bb G( \dot A^\perp,\delta_\perp A).
\end{align}

\bibliographystyle{JHEP}
\bibliography{Biblio_AsymptVarpi}

\end{document}